\documentclass[reprint, superscriptaddress, nofootinbib, amsmath, amssymb, aps, prd]{revtex4-2}

\usepackage{graphicx}
\usepackage{dcolumn}
\usepackage{bm}
\usepackage{hyperref}
\usepackage{color}
\usepackage{multirow}
\usepackage{comment}

\newcommand{\planck}[0]{\textit{Planck}}

\newcommand{\wmap}[0]{\textit{WMAP}}

\begin{document}

\title{Cosmic Birefringence from the Atacama Cosmology Telescope Data Release 6}

\author{Patricia Diego-Palazuelos}
\email{diegop@MPA-Garching.MPG.DE}
\affiliation{
Max-Planck-Institut f\"ur Astrophysik, Karl-Schwarzschild-Str. 1, D-85748 Garching, Germany
}

\author{Eiichiro Komatsu}
\affiliation{
Max-Planck-Institut f\"ur Astrophysik, Karl-Schwarzschild-Str. 1, D-85748 Garching, Germany
}
\affiliation{
Ludwig-Maximilians-Universit\"at M\"unchen, Schellingstr. 4, 80799 M\"unchen, Germany
}
\affiliation{Kavli Institute for the Physics and Mathematics of the Universe (Kavli IPMU, WPI), Todai Institutes for Advanced Study, The University of Tokyo, Kashiwa 277-8583, Japan}

\date{\today}

\begin{abstract}
The polarized light of the cosmic microwave background is sensitive to new physics that violates parity symmetry. For example, the interaction of photons with the fields of elusive dark matter and dark energy could cause a uniform rotation of the plane of linear polarization across the sky, an effect known as cosmic birefringence. We extract the cosmological rotation angle, $\beta$, using Bayesian analysis of parity-violating correlations, $EB$ and $TB$, of polarization data from the Atacama Cosmology Telescope (ACT) Data Release 6. We use prior probabilities for instrumental miscalibration angles derived from the optics model for the ACT telescope and instruments, and marginalize over a residual intensity-to-polarization leakage. We measure $\beta = 0.215^\circ\pm 0.074^\circ$ (68\% confidence level), which excludes $\beta=0$ with a statistical significance of $2.9\sigma$. Although there remain systematics in the ACT data that are not understood and do not allow us to draw strong cosmological conclusions, this result is consistent with previous independent results from the \wmap\ and \planck\ missions. It is suggestive that independent data sets and analyses using different methodologies have yielded the same sign and comparable magnitudes for $\beta$.
\end{abstract}

\maketitle

\section{\label{sec:intro}Introduction}

The most pressing questions in modern cosmology are the physical nature of dark matter and dark energy~\cite{Weinberg:2008zzc}. Despite the efforts of the research community over the past decades, we still do not know their precise nature. Is there information that we have not yet used? A violation of parity symmetry, the symmetry of a physical system under inversion of spatial coordinates, may provide new information \cite{Komatsu:2022nvu}. Since parity symmetry is broken in the weak interaction~\cite{Lee:1956qn, Wu:1957}, it is plausible that the physics behind them also breaks this symmetry. 

Parity-violating interactions of photons with the fields of dark matter and dark energy could cause a uniform rotation of the plane of linear polarization across the sky, an effect known as ``cosmic birefringence''~\cite{Carroll:1989vb, Carroll:1991zs, Harari:1992ea}. The isotropic rotation of the polarization plane by an angle $\beta$ breaks the parity symmetry because the rotation can only be clockwise ($\beta>0$) or counterclockwise ($\beta<0$) over the entire sky. 

The signal of cosmological parity violation has a clear signature in the linear polarization of the cosmic microwave background (CMB)~\cite{Lue:1998mq}. The observed polarization field can be decomposed into eigenstates of parity, $E$ and $B$ modes \cite{Kamionkowski:1996ks, Zaldarriaga:1996xe}. As they have opposite parity, their cross-correlation is sensitive to parity. The correlation functions of temperature ($T$) and polarization fields (or the power spectra, $C_\ell$, in spherical harmonics space with angular wave number $\ell$) contain four parity-even $TT$, $TE$, $EE$ and $BB$ power spectra, and two parity-odd $TB$ and $EB$ spectra. The uniform and frequency-independent rotation by an angle $\beta$ mixes the $E$ and $B$ modes, resulting in $C_\ell^{TB,\mathrm{o}}=\sin(2\beta)C_\ell^{TE}$ \cite{Lue:1998mq} and $C_\ell^{EB,\mathrm{o}}=\frac12 \sin(4\beta)(C_\ell^{EE}-C_\ell^{BB})$ \cite{Feng:2004mq}. Here, the superscript ``o'' denotes the observed value, whereas the power spectra on the right side are those before rotation. 

However, the observed rotation may also be due to instrumental systematics. The non-idealities of the optics used for observations artificially rotate the plane of linear polarization by another angle $\alpha$, which may be misinterpreted as a signal of $\beta$ if not corrected~\cite{Wu:2008qb, Miller:2009pt}. Measurements can only constrain the sum of the two angles, $\alpha+\beta$, unless we have information on $\alpha$.

In 2019, a solution to this problem was proposed~\cite{Minami:2019ruj,Minami:2020xfg, MinamiKomatsu:2020}. If the observed parity violation is of cosmological origin and is caused by a slowly evolving scalar field, the strength of the signal depends on the photon's path length. Therefore, we can break the degeneracy between $\alpha$ and $\beta$ by using polarized sources at very different locations in the past light cone, e.g., the polarized emission from gas in the interstellar medium of our Galaxy, emitted only $10^4$ light years away. The Galactic polarization is rotated only by $\alpha$, while the polarization of CMB photons, which have traveled 14 billion years, is rotated by $\alpha+\beta$.

Using the Galactic polarization to determine $\alpha$, tantalizing hints of cosmic birefringence have been reported from the analysis of the High-Frequency Instrument of ESA's \planck\ mission~\cite{Minami:2020odp,Diego-Palazuelos:2022dsq} and the joint analysis with the Low-Frequency Instrument~\cite{Eskilt:2022wav} and NASA's \wmap\ mission~\cite{Eskilt:2022cff}, $\beta=0.342^\circ{}^{+0.094^\circ}_{-0.091^\circ}$, which excludes $\beta=0$ with a $3.6 \sigma$ statistical significance. Throughout this paper, we quote uncertainties at the 68\,\% CL.

Further studies have been conducted to investigate various sources of systematics in this measurement, from instrumental systematics of \wmap\ and \planck\ \cite{Diego-Palazuelos:2022cnh, Cosmoglobe:2023pgf} to the possible $EB$ correlation intrinsic to Galactic emission \cite{Diego-Palazuelos:2022dsq, Diego-Palazuelos:2022cnh, Clark:2021kze, Hervias-Caimapo:2024ili}. Although none were able to explain the observed signal, systematics due to ``unknown unknowns'' cannot be ruled out until independent observations are made by other experiments. 

The Atacama Cosmology Telescope (ACT) collaboration has reported measurements of the $TB$ and $EB$ power spectra from the
Data Release 6 (DR6) \cite{ACT:2025fju}. They determined $\alpha$ using a model for the optics of the telescope and instruments \cite{Murphy:2024fna} without relying on the Galactic polarization. They found a rotation angle of $0.20^\circ\pm 0.08^\circ$ from the $EB$ correlation. 

However, Ref.~\cite{ACT:2025fju} did not explicitly estimate $\beta$. Instead, they used the $EB$ power spectrum to derive rotation angles for different detectors, $\psi_i$, and obtained the average angle, $\langle\psi_i\rangle\simeq 0.2^\circ$, with an uncertainty of $0.08^\circ$, including the calibration error estimated from the model for the ACT optics \cite{Murphy:2024fna}. This is a frequentist approach. In this paper, we use a Bayesian analysis to jointly estimate $\beta$ and $\alpha_i$ with prior probabilities of $\alpha_i$ derived from the optics model. We also include the $TB$ correlation and an additional channel at 220~GHz not used by the ACT team to derive angles. We further check the robustness of our results by marginalizing over a residual intensity-to-polarization ($I\to P$) leakage. 

\section{\label{sec:data}The ACT DR6}

We use the official cross array-band power spectra and covariance matrices published with the ACT DR6\footnote{\url{https://act.princeton.edu/act-dr6-data-products}}. See Refs.~\cite{ACT:2025xdm, ACT:2025fju, Atkins:2024jlo} for detailed descriptions of DR6 maps, power spectra, and covariance matrices. Here, we briefly summarize the key aspects relevant to our analysis.

The DR6 comprises data collected between 2017 and 2022 and covers 19,000\,deg$^2$ of sky with a median combined depth of $10\,\mu$K\,arcmin and an angular resolution of an arcminute~\cite{ACT:2025xdm}. 
The observations include the mid-frequency polarization arrays (PA) PA5 and PA6, operating at frequencies f090 (77–112\,GHz) and f150 (124–172\,GHz), and the high-frequency array PA4, operating at f150 and f220 (182–277\,GHz).
The ACT team excluded PA4\,f150 temperature and polarization data from cosmological analyses due to the failure of multiple null tests. They also excluded the PA4\,f220 polarization due to its limited constraining power compared to other array-bands, but we include it in our analysis.

From the total footprint, only 10,000\,deg$^2$ are used to calculate the power spectra after masking regions with high Galactic emission and extragalactic point sources~\cite{ACT:2025fju}. The power spectra are corrected for the transfer function from the ground pickup filter, $I\to P$ leakage~\cite{Duivenvoorden}, and the aberration introduced by our motion relative to the last scattering surface. In addition to encoding the inhomogeneous noise properties and complex correlation patterns arising from the scanning strategy and ground pickup filter, the covariance matrix includes contributions due to uncertainties in the beam characterization and the beam leakage correction, as well as non-Gaussian corrections from lensing and foreground emission~\cite{Atkins:2024jlo}. The resulting temperature and polarization data are signal-dominated up to multipoles $\ell = 2800$ and 1700, respectively.

\begin{figure}
\centering
\includegraphics[width=0.9\linewidth]{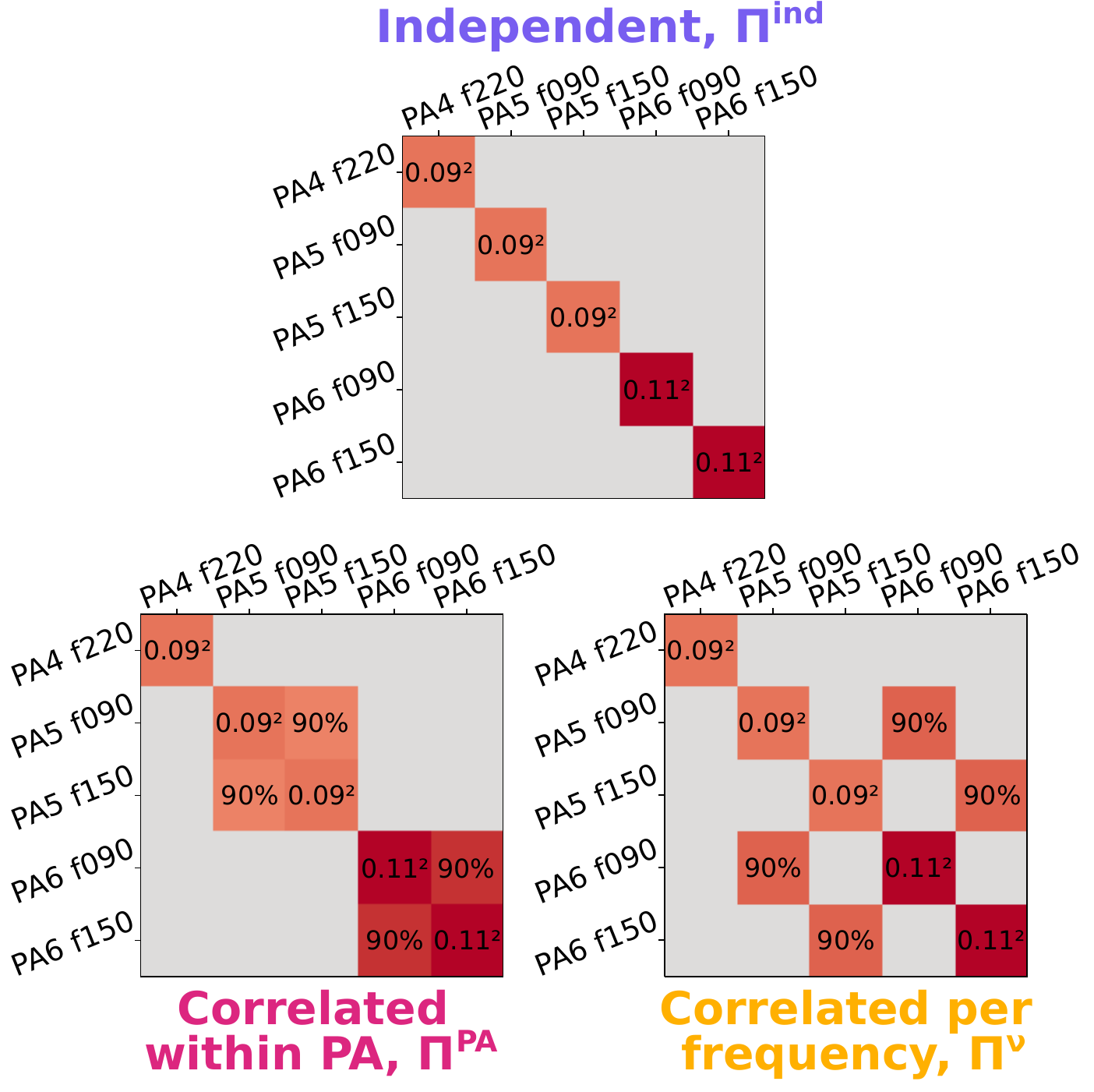}
\caption{\label{fig:priors}
Covariance matrix of the Gaussian priors on miscalibration angles used in our analysis. The diagonal elements are in units of degrees squared, while the off-diagonal elements indicate the correlation coefficient, $\rho$, between array-bands.}
\end{figure}

Ref.~\cite{Murphy:2024fna} quantified systematic uncertainties in determining the polarization angles (miscalibration angles, $\alpha_i$) of the $i$th array-band with respect to the sky coordinates. They account for possible displacements in lens positions and orientations, and anti-reflection coating thickness and refractive indices using polarization-sensitive ray tracing software. The standard deviations of systematic uncertainties per PA are $\sigma_{\alpha_\mathrm{PA4}}=0.09^\circ$, $\sigma_{\alpha_\mathrm{PA5}}=0.09^\circ$, and $\sigma_{\alpha_\mathrm{PA6}}=0.11^\circ$, with polarization angles rotating by less than $0.01^\circ$ within the passbands. The frequency dependence of the dichroic detectors' response or any subsequent data processing is not modeled.

In this paper, we explore three sets of Gaussian priors from this optical model: (1) ``Independent'' ($\Pi^\mathrm{ind}$), where $\alpha_i$ are independent of $i$ with the standard deviations given by $\sigma_{\alpha_\mathrm{PA4}}$, $\sigma_{\alpha_\mathrm{PA5}}$, and $\sigma_{\alpha_\mathrm{PA6}}$; (2) ``Correlated within PA'' ($\Pi^\mathrm{PA}$), where $\rho=90\%$ correlation between bands within the same PA is enforced; and (3) ``Correlated per frequency'' ($\Pi^\nu$), where $\rho=90\%$ correlation between bands observing at the same frequency is enforced. This last case is inspired by the hints of a frequency-dependent pattern seen in Figure 9 of Ref.~\cite{ACT:2025fju}, with f090 yielding lower rotation angles than f150 in both PA5 and PA6. Priors are visualized in Fig.~\ref{fig:priors}. The ACT team expects $\alpha_i$ within the same PA to be highly correlated as they are housed in the same optics tube, sharing all optical components. Therefore, the baseline configuration is $\Pi^\mathrm{PA}$, and the other configurations are used to assess the robustness of the results with respect to variations in the choice of priors.

\section{\label{sec:analysis}Analysis method}

Our analysis pipeline is based upon Refs.~\cite{Minami:2019ruj, Minami:2020xfg, MinamiKomatsu:2020}, extending the formalism to jointly analyze $EB$ and $TB$ correlations and including priors on $\alpha_i$. The code to reproduce the results of this paper is publicly available.\footnote{\url{https://github.com/pdp79/act_dr6_analysis}}

It is not possible to distinguish between $\beta$ and  $\alpha_i$, if we rely only on the CMB $C_\ell^{EB}$ and $C_\ell^{TB}$~\cite{Wu:2008qb,Miller:2009pt,Komatsu:2010fb,Keating:2012ge}. Thus, additional information is needed to break the degeneracy between them. We use the model for the ACT optics to constrain $\alpha_i$.

We use five polarized array-bands from ACT DR6, $i\in\{\mathrm{PA}4\,\mathrm{f}220, \mathrm{PA}5\,\mathrm{f}090, \mathrm{PA}5\,\mathrm{f}150, \mathrm{PA}6\,\mathrm{f}090, \mathrm{PA}6\,\mathrm{f}150\}$, assigning a different $\alpha_i$ to each. Unlike in previous \wmap\ and \planck\ analyses~\cite{Minami:2020odp, Diego-Palazuelos:2022dsq, Eskilt:2022wav, Eskilt:2022cff, Cosmoglobe:2023pgf}, we do not need to exclude the auto-power spectra of the same band, e.g., PA5\,f090$\times$PA5\,f090, to avoid potential contamination from correlated noise. This is because the DR6 power spectra are computed by cross-correlating independent splits of observations~\cite{ACT:2025fju}. 
We adhere to the binning scheme adopted by the ACT team (see Ref.~\cite{ACT:2025fju} for details), characterized by a dynamical width that increases as the signal-to-noise decreases. The typical width is $\Delta\ell=50$ for the approximate $\ell\sim 120-2000$ range, rising to several hundred $\Delta\ell$ for lower and higher multipoles. For each array-band, we use the multipole bins of mean multipole, $b$, within the range $b_\mathrm{min}\leq b \leq b_\mathrm{max}$ shown in Table~\ref{tab:our multipole cut}. This choice is inspired by the baseline multipole cut defined in Table~1 of Ref.~\cite{ACT:2025fju}, additionally setting $b_\mathrm{min}=1000.5$ for PA4\,f220. 
For the rest of this paper, we work with $D_\ell^{XY}\equiv \ell(\ell+1)C_\ell^{XY}/(2\pi)$, where $X$ and $Y$ are any of $(T,E,B)$.

\begin{table}
	\caption{\label{tab:our multipole cut}%
		Minimum and maximum bin center of ACT DR6 power spectra used in our analysis. For each array-band, this amounts to a total number of bins $N_b$.}
	\begin{ruledtabular}
		\begin{tabular}{cccc}
			Array-band & $b_\mathrm{min}$ & $b_\mathrm{max}$ & $N_b$ \\
			\colrule
            PA4\,f220 & 1000.5 & 7525.5 & 39 \\
            PA5\,f090 & 1000.5 & 7525.5 & 39 \\
            PA5\,f150 & 800.5 & 7525.5 & 43 \\
            PA6\,f090 & 1000.5 & 7525.5 & 39 \\
            PA6\,f150 & 600.5 & 7525.5 & 47 \\
		\end{tabular}
	\end{ruledtabular}
\end{table}

We neglect the impact of the dust $TB$ and $EB$ correlations~\cite{Huffenberger:2019mjx, Clark:2021kze}. The justification for this decision is twofold. First, at the angular scales used in our analysis ($\ell \geq 600$), ACT's polarization power spectra are dominated by CMB and noise after applying the extension to the \planck\ 70\% Galactic mask presented in section C.2 of Ref.~\cite{ACT:2025fju}. Second, our $\beta$ estimates are not as sensitive to parity-violating dust signals as those in Ref.~\cite{Diego-Palazuelos:2022dsq} since we do not rely on foreground emission to break the $\alpha_i+\beta$ degeneracy.

We define the posterior, $\mathcal{P}$, as
\begin{equation}
-2\ln\mathcal{P} = \sum_{\substack{b,b'=\\\mathrm{max}\{b_\mathrm{min},b'_\mathrm{min}\}}}^{b_\mathrm{max}} \vec{v}^{\sf T}_b\mathbf{M}^{-1}_{b,b'}\vec{v}_{b'} +\ln|\mathbf{M}_{b,b'}|+\vec{\alpha}^{\sf T}\Pi^{-1}\vec{\alpha}\,,
\end{equation}
and use the iterative semi-analytical formalism described in Refs.~\cite{delaHoz:2021vfx,Diego-Palazuelos:2022cnh} to calculate the posterior distributions for $\beta$ and $\alpha_i$ in the small-angle approximation, $|\alpha_i|\ll 1$ and $|\beta|\ll1$.  
Here, $b$ is the index for bins, and $\vec{v}_b\equiv \mathbf{A}\vec{D}_b^\mathrm{o}-\mathbf{B}\vec{D}_b^\mathrm{CMB}$ is the data vector built from the one-dimensional arrays $\vec{D}_b^\mathrm{o}=(D_b^{E_iE_j,\mathrm{o}}, D_b^{B_iB_j,\mathrm{o}}, D_b^{E_iB_j,\mathrm{o}}, D_b^{T_iE_j,\mathrm{o}}, D_b^{T_iB_j,\mathrm{o}})^{\sf T}$ and $\vec{D}_b^\mathrm{CMB}=(D_b^{EE,\mathrm{CMB}}, D_b^{BB,\mathrm{CMB}}, D_b^{TE,\mathrm{CMB}})^{\sf T}$, $\mathbf{M}_{b,b'} = \mathbf{A}\mathrm{Cov}(\vec{D}_b^\mathrm{o}, \vec{D}_{b'}^{\mathrm{o}\sf T})\mathbf{A}^{\sf T}$ the covariance matrix, and $\Pi$ a Gaussian prior on the $\vec{\alpha}=(\alpha_i)^{\sf T}$ with zero mean. $\mathbf{A}$ and $\mathbf{B}$ are block-diagonal matrices given by
\begin{align}
\nonumber
\mathbf{A} =
\mathrm{diag}&\left(-\frac{\sin(4\alpha_j)}{\cos(4\alpha_i)+\cos(4\alpha_j)}, \frac{\sin(4\alpha_i)}{\cos(4\alpha_i)+\cos(4\alpha_j)}, \right. \\
&\left. 1, -\tan(2\alpha_j), 1\right)\,,\\
\nonumber
\mathbf{B} = \mathrm{diag}&\left(\frac{\sin(4\beta)}{2\cos(2\alpha_i+2\alpha_j)}, -\frac{\sin(4\beta)}{2\cos(2\alpha_i+2\alpha_j)}, \right.\\
& \left. \frac{\sin(2\beta)}{2\cos(2\alpha_j)}\right)\,. 
\end{align}
We use the beam-deconvolved and binned array-band power spectra, $D_b^{X_iY_j,\mathrm{o}}$, and the non-diagonal covariance matrices including bin-to-bin correlations, $\mathrm{Cov}(D_b^{X_iY_j,\mathrm{o}},D_{b'}^{W_pZ_p,\mathrm{o}})$, published with the ACT DR6 to build $\vec{D}_b^\mathrm{o}$ and $\mathbf{M}_{b,b'}$. $\vec{D}_b^\mathrm{CMB}$ is built from the binned theoretical $\Lambda$ Cold Dark Matter ($\Lambda$CDM) CMB spectra that best fit the ACT DR6~\cite{ACT:2025fju}. For cross-spectra between array-bands of different $b_\mathrm{min}$, we keep the bins within the higher $b_\mathrm{min}$. For example, within PA6 we use $b\geq 600.5$ for the f150$\times$f150 spectrum, but $b\geq1000.5$ for the f090$\times$f090 and f090$\times$f150 spectra. Thus, the combination of the 5 bands provides $995\times 2$ data points when using $EB$ and $TB$.

\section{\label{sec:results}Results}

We first validate our pipeline against the results obtained by the ACT team by fitting only for $\psi_i=\alpha_i+\beta$.
We exclude PA4\,f220 and use only $EB$ information within the baseline multipole cut.
We find $\psi_i = 0.108^\circ\pm0.064^\circ$, $0.316^\circ\pm0.059^\circ$, $0.118^\circ\pm0.074^\circ$, and $0.185^\circ\pm0.060^\circ$ for $i=\mathrm{PA}5\,\mathrm{f}090$, $\mathrm{PA}5\,\mathrm{f}150$, $\mathrm{PA}6\,\mathrm{f}090$, and $\mathrm{PA}6\,\mathrm{f}150$, respectively, in agreement with the values reported in the bottom left panel of Fig.~9 in Ref.~\cite{ACT:2025fju}. The combination of such $\psi_i$ through the weighted average, $\langle\psi_i\rangle = \left[\sum_{ik}\psi_i(C^{-1})_{ik} \right]/\sum_{jk}(C^{-1})_{jk}$, with associated uncertainty $\sigma = \left[\sum_{jk}(C^{-1})_{jk}\right]^{-1/2}$, yields a global rotation angle of $\langle\psi_i\rangle = 0.192^\circ\pm0.032^\circ$, in agreement with $0.20^\circ\pm0.03^\circ$ reported in Ref.~\cite{ACT:2025fju} before adding systematic errors in quadrature.

As pointed out by the ACT team, we have reproduced the discrepancy between the angles within the PA5, $\psi_\mathrm{PA5\,f150}-\psi_\mathrm{PA5\,f090}=0.21^\circ\pm 0.09^\circ$, with modest statistical significance. They report that ``This is unexpected given that these arrays are in the same optics tube, so at present we lack an instrument-based model to explain this discrepancy'' \cite{ACT:2025fju}. As we demonstrate in this paper, the statistical significance and interpretation of this discrepancy depend on the choice of prior distribution for $\alpha_i$ based on the ACT optics model.

\begin{table}
	\caption{\label{tab:result alpha beta}%
		Cosmic birefringence and miscalibration angles in units of degrees with $1\,\sigma~(68\,\%)$ uncertainties derived from ACT's DR6 $EB$ and $TB$ power spectra when using different priors. The $\chi^2$ has 1984 degrees of freedom.}
	\begin{ruledtabular}
		\begin{tabular}{cccc}
			\multirow{2}{*}{Prior} & \multirow{2}{*}{Independent} & 90\% correlation  & 90\% correlation  \\
             &  &  within PA &  per frequency \\
			\colrule
			$\beta$                     & $\phantom{-}0.204 \pm 0.058$ & $\phantom{-}0.212 \pm 0.072$ & $\phantom{-}0.212 \pm 0.072$ \\
			$\alpha_\mathrm{PA4\,f220}$ & $\phantom{-}0.023 \pm 0.085$ & $\phantom{-}0.023 \pm 0.085$ & $\phantom{-}0.022 \pm 0.085$ \\
			$\alpha_\mathrm{PA5\,f090}$ & $-0.063 \pm 0.064$ & $-0.012 \pm 0.072$ & $-0.065 \pm 0.068$ \\
			$\alpha_\mathrm{PA5\,f150}$ & $\phantom{-}0.080 \pm 0.063$ & $\phantom{-}0.026 \pm 0.071$ & $\phantom{-}0.044 \pm 0.068$ \\
			$\alpha_\mathrm{PA6\,f090}$ & $-0.038 \pm 0.072$ & $-0.035 \pm 0.077$ & $-0.069 \pm 0.080$ \\
			$\alpha_\mathrm{PA6\,f150}$ & $-0.008 \pm 0.068$ & $-0.026 \pm 0.076$ & $\phantom{-}0.021 \pm 0.079$ \\
            \colrule
            $\chi^2$ & 2031 & 2034  & 2032 \\
            PTE      & 22.7\,\% & 21.1\,\% & 22.1\,\%\\
		\end{tabular}
	\end{ruledtabular}
\end{table}

\begin{figure}
\centering
\includegraphics[width=0.9\linewidth]{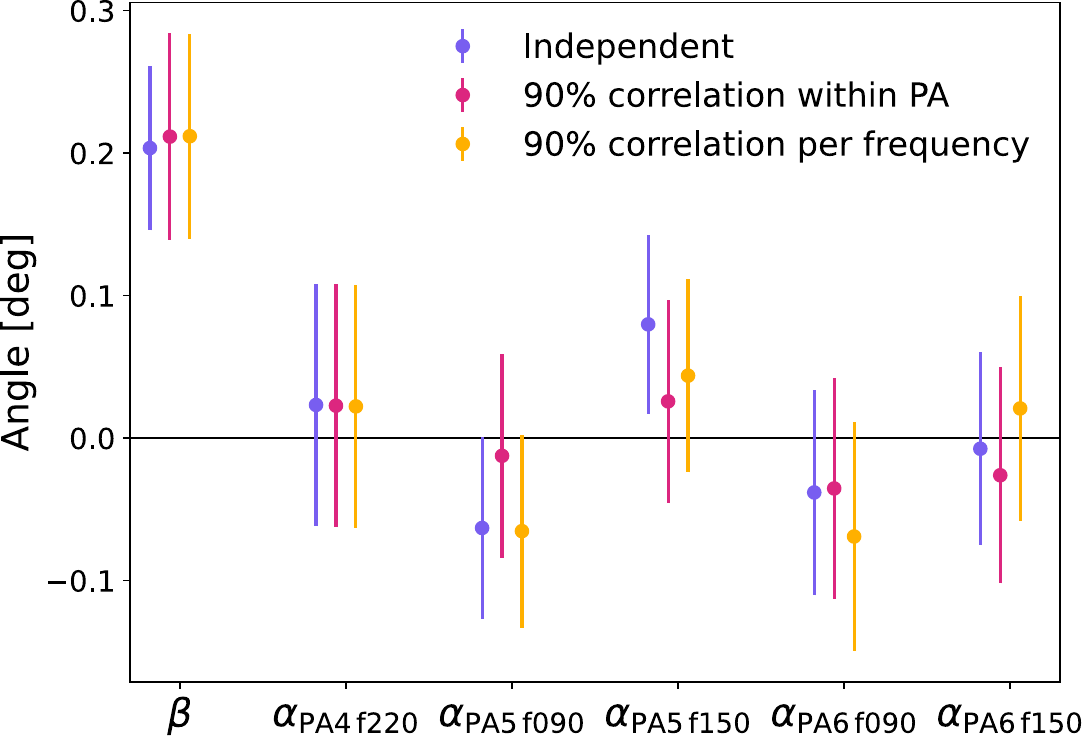}
\caption{\label{fig:angles}
Cosmic birefringence and miscalibration angles derived from ACT's DR6 $EB$ and $TB$ power spectra when using different priors. }
\end{figure}

We now perform a Bayesian analysis using prior information on the instrument model to simultaneously determine $\beta$ and $\alpha_i$. 
Table~\ref{tab:result alpha beta} and Fig.~\ref{fig:angles} show the $\beta$ and $\alpha_i$  from the joint analysis of $C_\ell^{EB}$ and $C_\ell^{TB}$ including PA4\,f220. For all priors on $\alpha_i$, we find $\beta\simeq 0.2^\circ$. We consider all bin-to-bin off-diagonal correlations in the covariance matrix for our baseline results. Dropping off-diagonal terms leads to $\leq0.01^\circ$ shifts on the best-fit values.
Compared to the prior knowledge of the optics model, the joint analysis of $EB$ and $TB$ spectra tightens the constraints on $\alpha_i$ for PA5 and PA6 by 20 to 37\%. However, constraints on PA4\,f220 are mostly dominated by the knowledge of instrumental miscalibrations, given the noisier nature of this band~\cite{ACT:2025xdm}. As such, including this band only provides a 1.05 to 3.53\,\% improvement in signal-to-noise.

Our Bayesian framework propagates the systematic uncertainties $\alpha_i$ to the final $\beta$ through the use of priors, leading to an increase in the uncertainty from $\sigma_\beta=0.032^\circ$ to $0.058^\circ$ for $\Pi^\mathrm{ind}$, and to $0.072^\circ$ for $\Pi^\mathrm{PA}$ and $\Pi^\nu$. We report the values of $\chi^2= \sum_{b,b'=\mathrm{max}\{b_\mathrm{min},b'_\mathrm{min}\}}^{b_\mathrm{max}}  \vec{v}^{\sf T}_b\mathbf{M}^{-1}_{b,b'}\vec{v}_{b'}$ and the probability-to-exceed (PTE) in Table~\ref{tab:result alpha beta}.

The recovered $\alpha_i$ are consistent between array-bands and choices of prior. 
We find that the $2.0\,\sigma$ discrepancy between $\alpha_\mathrm{PA5\,f090}$ and $\alpha_\mathrm{PA5\,f150}$ seen for $\Pi^\mathrm{ind}$ reduces to $1.1\,\sigma$ and $1.9\,\sigma$ for $\Pi^\mathrm{PA}$ and $\Pi^\nu$, respectively, at the expense of a slightly larger $\chi^2$. This occurs because the baseline prior from the optics model prefers similar $\alpha_i$ within the same PA, bringing the best-fitting $\alpha_\mathrm{PA5\,f090}$ and $\alpha_\mathrm{PA5\,f150}$ closer together. 

At the heart of Bayesian analysis is the idea that the interpretation of the results depends on the prior probability of the instrument model. It is the degree of belief in our interpretation given our knowledge of the instrument. If the data conclusively show that $\alpha_\mathrm{PA5\,f090}$ and $\alpha_\mathrm{PA5\,f150}$ are discrepant, the data will overcome the prior. However, despite the mild preference for divergent angles, interpretations suggesting that these channels are not discrepant are also consistent with the data, since $\chi^2$ is not very different for different priors.

As discussed above, the discrepancy between $\alpha_\mathrm{PA5\,f090}$ and $\alpha_\mathrm{PA5\,f150}$ made Ref.~\cite{ACT:2025fju} raise concerns about their instrument model.
Although our results provide a Bayesian framework that can accommodate this discrepancy, we do not find a significant $\Delta\chi^2$ indicating a statistical preference for one prior over another.
Similar results are obtained for priors with higher correlation coefficients $\rho$.

For completeness, we also calculate the rotation angles using only $TB$ for the baseline $\Pi^\mathrm{PA}$ prior. We find $\beta=0.254^\circ\pm0.131^\circ$ and $\alpha_i = 0.001^\circ\pm0.089^\circ$, $0.012^\circ\pm0.085^\circ$, $0.016^\circ\pm0.085^\circ$, $-0.013^\circ\pm0.101^\circ$, and $-0.008^\circ\pm0.100^\circ$  for $i=\mathrm{PA4\,f220}$, PA5\,f090, PA5\,f150, PA6\,f090, and PA6\,f150, respectively. As expected, $TB$ provides a lower signal-to-noise than $EB$~\cite{Aghanim:2016fhp,Sullivan:2025btc}, thus leading to results that are dominated by the precision of the optics model. 
Indeed, the inclusion of $TB$ information only tightens constraints by 0.23 to 1.24\,\% compared to an $EB$-only analysis. Still, the consistent $\beta\simeq 0.2^\circ$ found in $TB$ shows that our $\beta$ estimate is not significantly influenced by additional systematics beyond $\alpha_i$ miscalibrations. Similarly, the stable $\beta\simeq 0.2^\circ$ value found when including PA4\,f220 shows that our analysis does not suffer from significant dust contamination.

\begin{figure}
\centering
\includegraphics[width=1\linewidth]{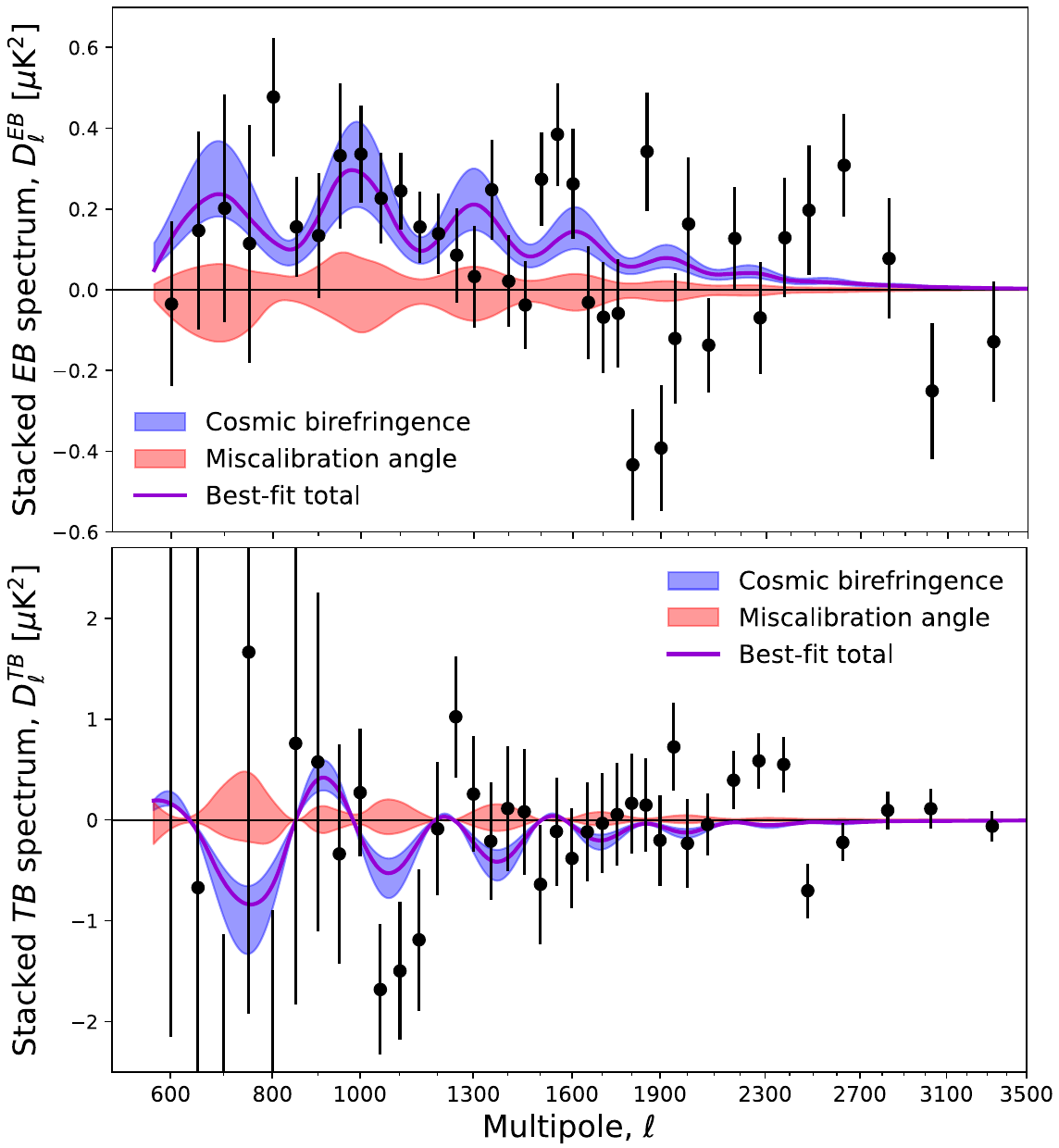}
\caption{\label{fig:stacked_specs}
Stacked $EB$ (upper) and $TB$ (lower) power spectra observed by ACT DR6's PA4\,f220, PA5\,f090, PA5\,f150, PA6\,f090, and PA6\,f150 in the baseline multipole cut. The red and blue bands show, respectively, the $1\,\sigma$ confidence contours from the polarization angle miscalibration and cosmic birefringence contributions to our joint $EB$ and $TB$ fit with $\Pi^\mathrm{PA}$. The purple line shows the total best-fitting model.}
\end{figure}

Finally, Fig.~\ref{fig:stacked_specs} shows the stacked $EB$ (upper panel) and $TB$ (lower panel) power spectra observed by ACT DR6 and the $\beta$ (blue) and $\alpha_i$ (red) contributions obtained in our joint fit to $EB$ and $TB$. Following Ref.~\cite{Eskilt:2022cff}, we calculate the inverse-variance weighted average across frequencies from $\mathrm{E}[D_b^{XY}] = \sum_{ijpq}(\mathbf{C}^{ijpq}_{XY,b})^{-1}D_{b}^{X_pY_q,\mathrm{o}}/\sum_{ijpq}(\mathbf{C}^{ijpq}_{XY,b})^{-1}$ with variance $\mathrm{Var}[D_b^{XY}] = 1/\sum_{ijpq}(\mathbf{C}^{ijpq}_{XY,b})^{-1}$, where $\mathbf{C}^{ijpq}_{XY,b}=\mathrm{Cov}(D_b^{X_iY_j,\mathrm{o}},D_{b}^{X_pY_q,\mathrm{o}})$.

\section{\label{sec:systematics}Impact of residual \texorpdfstring{$I\to P$}{I->P} leakage}

Beyond the systematics on polarization angle calibration, spurious $TB$ and $EB$ correlations might also appear from an imperfect knowledge of the instrument's beam~\cite{Miller:2009pt}. We test the robustness of our results against a residual $I\to P$ leakage that might remain after ACT's processing~\cite{ACT:2025xdm, ACT:2025fju, Duivenvoorden}.

We extend our model of the spherical harmonics coefficients of the temperature and $E$ and $B$ modes, $\vec{X}_{\ell m} = ( T_{\ell m}, E_{\ell m}, B_{\ell m})^{\sf T}$, observed at every array-band to $\vec{X}^{i,\mathrm{o}}_{\ell m}=\mathbf{F}\vec{X}_{\ell m}$, which encodes the impact of $\beta$, $\alpha_i$, and $I\to P$ leakage through the mixing matrix,
\begin{equation}
\mathbf{F}
=
\left(\begin{array}{ccc}
1 & 0 & 0 \\
\gamma_i & \cos(2\alpha_i+2\beta) & -\sin(2\alpha_i+2\beta)\\
\gamma_i & \sin(2\alpha_i+2\beta) & \phantom{-}\cos(2\alpha_i+2\beta)
\end{array}\right)\,,
\end{equation}
where $\gamma_i$ characterizes the residual $I\to P$ leakage based on a simplification of the data model adopted in Ref.~\cite{ACT:2025fju}, where $\gamma_i$ coefficients were allowed to vary across angular scales and between $TE$ and $TB$, $\gamma_{i,\ell}^{TE}$ and $\gamma_{i,\ell}^{TB}$. Such mixing of temperature and polarization signals introduces couplings of the $\beta$, $\alpha_i$, and $\gamma_i$ that greatly complicate our data model. However, a Taylor expansion up to first order in $\beta$, $\alpha_i$, and $\gamma_i$ provides a good approximation.
We then only need to modify $\mathbf{B}$ to $\mathrm{diag}(2\beta,-2\beta~,\beta+\gamma_j,\gamma_j)$ and $\vec{D}_b^\mathrm{CMB}$ to $(D_b^{EE,\mathrm{CMB}},D_b^{BB,\mathrm{CMB}},D_b^{TE,\mathrm{CMB}},D_b^{TT,\mathrm{CMB}})^{\sf T}$ in our posterior to simultaneously fit for all three effects. We keep the covariance matrix unchanged.

Table~\ref{tab:result alpha beta gamma} shows the $\beta$, $\alpha_i$, and $\gamma_i$ from the joint fit to the $EB$ and $TB$ power spectra for different priors. The full corner plot displaying all correlations between parameters is shown in Appendix~\ref{sec:appendix}. The small $|\gamma_i|\leq0.08\,\%$ values obtained for residual $I\to P$ leakage fall well below the original 0.3\,\% leakage correction applied to the ACT DR6 maps~\cite{ACT:2025fju}. Marginalizing over the residual $I\to P$ leakage produces only small shifts in our posteriors, shifting $\alpha_i$ by $\leq0.21\sigma$ and $\beta$ by $\leq0.07\sigma$. We conclude that the $I\to P$ leakage has no significant impact on our results. 

\begin{table}
	\caption{\label{tab:result alpha beta gamma}%
		Cosmic birefringence, miscalibration angles, and amplitude of the residual $I\to P$ leakage with $1\,\sigma~(68\,\%)$ uncertainties derived from the joint fit of ACT's DR6 $EB$ and $TB$ power spectra when using different priors. Angles are in units of degrees while $\gamma$ coefficients are adimensional. The $\chi^2$ has 1979 degrees of freedom.}
	\begin{ruledtabular}
		\begin{tabular}{cccc}
			\multirow{2}{*}{Prior} & \multirow{2}{*}{Independent} & 90\% correlation  & 90\% correlation  \\
             &  &  within PA &  per frequency \\
			\colrule
			$\beta$                     & $\phantom{-}0.208 \pm 0.059$ & $\phantom{-}0.215 \pm 0.074$ & $\phantom{-}0.216 \pm 0.073$ \\
			$\alpha_\mathrm{PA4\,f220}$ & $\phantom{-}0.025 \pm 0.085$ & $\phantom{-}0.025 \pm 0.085$ & $\phantom{-}0.024 \pm 0.085$\\
			$\alpha_\mathrm{PA5\,f090}$ & $-0.063 \pm 0.064$           & $-0.014 \pm 0.072$           & $-0.059 \pm 0.068$ \\
			$\alpha_\mathrm{PA5\,f150}$ & $\phantom{-}0.075 \pm 0.063$ & $\phantom{-}0.023 \pm 0.072$ & $\phantom{-}0.037 \pm 0.068$ \\
			$\alpha_\mathrm{PA6\,f090}$ & $-0.023 \pm 0.072$           & $-0.030 \pm 0.077$           & $-0.058 \pm 0.080$ \\
			$\alpha_\mathrm{PA6\,f150}$ & $-0.018 \pm 0.068$           & $-0.029 \pm 0.076$           & $\phantom{-}0.011 \pm 0.079$ \\
            $10^2\gamma_\mathrm{PA4\,f220}$ & $\phantom{-}0.050 \pm 0.152$ & $\phantom{-}0.051 \pm 0.152$ & $\phantom{-}0.051 \pm 0.152$ \\
            $10^2\gamma_\mathrm{PA5\,f090}$ & $\phantom{-}0.014 \pm 0.046$ & $\phantom{-}0.015 \pm 0.046$ & $\phantom{-}0.015 \pm 0.046$ \\
            $10^2\gamma_\mathrm{PA5\,f150}$ & $-0.011 \pm 0.042$           & $-0.010 \pm 0.042$           & $-0.010 \pm 0.042$ \\
            $10^2\gamma_\mathrm{PA6\,f090}$ & $\phantom{-}0.081 \pm 0.050$ & $\phantom{-}0.083 \pm 0.050$ & $\phantom{-}0.083 \pm 0.050$ \\
            $10^2\gamma_\mathrm{PA6\,f150}$ & $-0.054 \pm 0.046$           & $-0.053 \pm 0.046$           & $-0.053 \pm 0.046$ \\
            \colrule
            $\chi^2$ & 2027 & 2030 & 2029 \\
            PTE      & 22.1\,\% & 20.7\,\% & 21.4\,\%\\
		\end{tabular}
	\end{ruledtabular}
\end{table}

\section{\label{sec:conclusion}Conclusions}

In this paper, we searched for the signal of cosmic birefringence in the ACT DR6 data. We explicitly included the frequency-independent cosmological rotation angle, $\beta$, and the prior probabilities for the miscalibration angles, $\alpha_i$, derived from the ACT optics model of Ref.~\cite{Murphy:2024fna} in the likelihood for the $TB$ and $EB$ power spectra.
We measured $\beta = 0.215^\circ\pm 0.074^\circ$, which excludes $\beta=0$ with a statistical significance of $2.9\sigma$. Our result is compatible with the mean rotation $\langle\psi_i\rangle = 0.20^\circ \pm 0.08^\circ$ reported by the ACT team~\cite{ACT:2025fju}, but provides an alternative interpretation that discerns array-band angle miscalibrations from a cosmological rotation. 

This result is consistent with previous results from the \wmap\ and \planck\ missions~\cite{Minami:2020odp, Diego-Palazuelos:2022dsq, Eskilt:2022wav, Eskilt:2022cff, Cosmoglobe:2023pgf}. The joint analysis of the original \wmap\ 9-year data \cite{Hinshaw:2012aka} and the \planck\ Public Release 4 \cite{Akrami:2020bpw} gave $\beta=0.342^\circ{}^{+0.094^\circ}_{-0.091^\circ}$ \cite{Eskilt:2022cff} and the reanalysis using the Cosmoglobe-processed \wmap\ and \planck\ data gave $\beta=0.26^\circ\pm 0.10^\circ$ \cite{Cosmoglobe:2023pgf}. Assuming that the ACT and \wmap+\planck\ results are statistically independent due to a limited overlap in the multipole ranges used for the analysis, we find $\beta = 0.264^\circ\pm 0.058^\circ$ and $0.231^\circ\pm 0.059^\circ$, which excludes $\beta=0$ with a statistical significance of $4.6\sigma$ and $3.9\sigma$, respectively. However, there are still unresolved systematics in the ACT data, such as the discrepancy between the angles derived from the 90 and 150~GHz data in PA5 with moderate significance \cite{ACT:2025fju}, which do not yet allow us to draw strong cosmological conclusions.

This work does not represent an incremental improvement in the measurement of $\beta$. It is significant that independent data sets and analyses using different methodologies have yielded the same sign and comparable magnitudes for $\beta$. As the statistical significance of $\beta$ approaches $5\sigma$, controlling systematics is key.

To move forward, we need independent confirmation from experiments using artificial polarization sources as calibrators, such as BICEP3~\cite{BICEPKeck:2024cmk}, CLASS~\cite{Coppi:2025fmt}, and the Simons Observatory~\cite{Murata:2023heo}, which do not rely on models of optics or Galactic polarization. If confirmed, the discovery of cosmological parity violation is a clear sign of new physics beyond the standard model of elementary particles and fields and will have a revolutionary impact on cosmology and fundamental physics.

\begin{acknowledgments}
We thank A. J. Duivenvoorden for his help with the likelihood products. We also thank S. Choi, S. Clark, J. Dunkley, J.~C. Hill,  T. Louis, S. Næss, and L. Page for useful discussions, and J.~R. Eskilt for comments on the draft. Some of the results in this paper have been derived using the \texttt{numpy}~\cite{Harris:2020xlr}, \texttt{matplotlib}~\cite{Hunter:2007ouj}, and \texttt{GetDist}~\cite{Lewis:2019xzd} packages. This work was supported in part by JSPS KAKENHI Grant No.~JP20H05850 and JP20H05859, and the Deutsche Forschungsgemeinschaft (DFG, German Research Foundation) under Germany's Excellence Strategy - EXC-2094 - 390783311. This work has also received funding from the European Union’s Horizon 2020 research and innovation programme under the Marie Skłodowska-Curie Grant Agreement No. 101007633.
The Kavli IPMU is supported by the World Premier International Research Center Initiative (WPI), MEXT, Japan.
\end{acknowledgments}
\bibliographystyle{apsrev4-2}
\bibliography{references}

\appendix
\section{\label{sec:appendix} Fit to cosmic birefringence, polarization angle miscalibration, and residual $I\to P$ leakage}

\begin{figure*}
\centering
\includegraphics[width=1\linewidth]{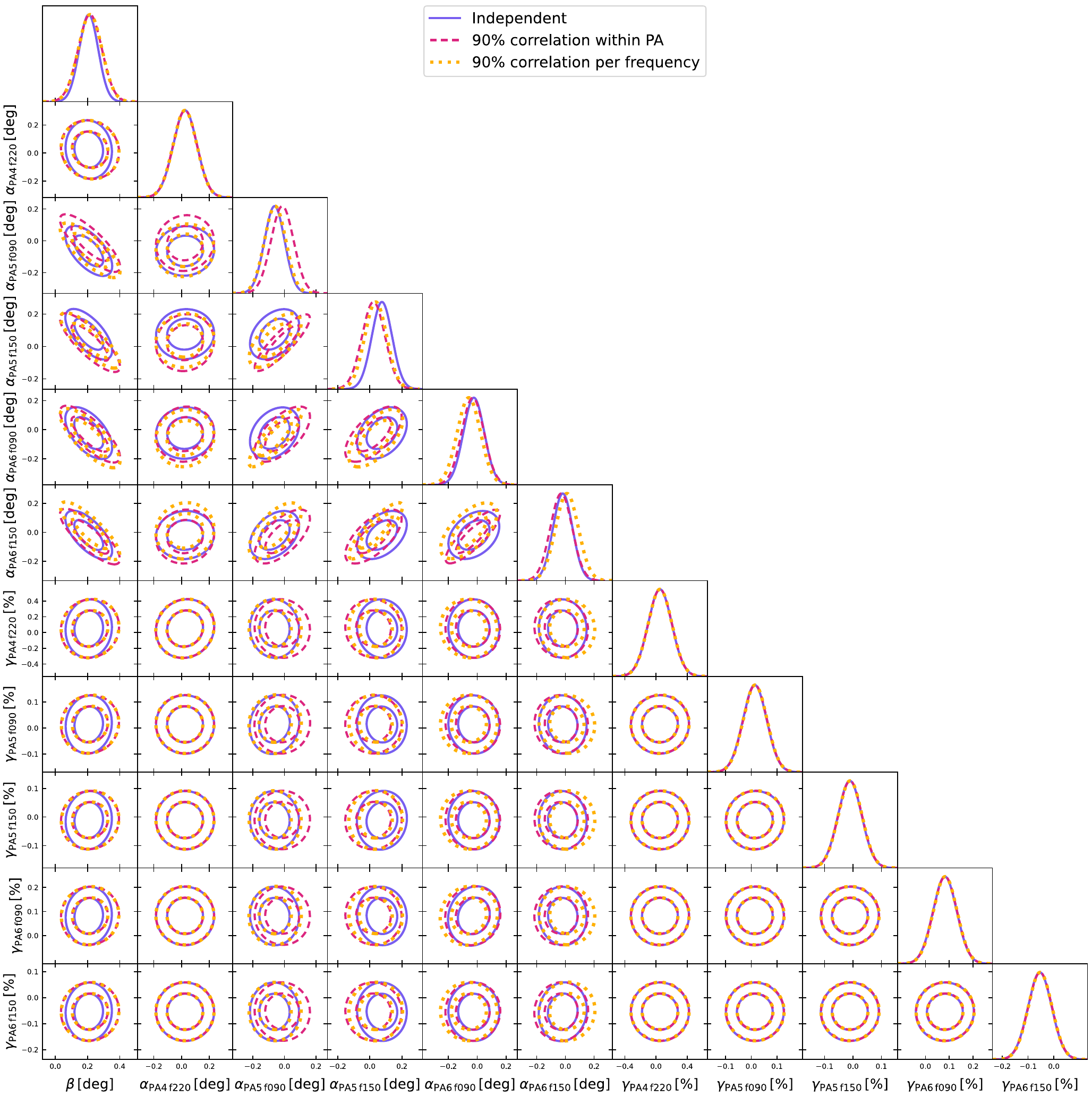}
\caption{\label{fig:corner plot}
Birefringence, miscalibration angles, and amplitudes of residual $I\to P$ leakage derived from ACT's DR6 $EB$ and $TB$ power spectra when using different priors.}
\end{figure*}

\end{document}